\documentstyle[aps,preprint,epsf]{revtex}
\def\a{\alpha}
\def\b{\beta}

\def\d{\delta}

\def\m{\mu}
\def\n{\nu}

\def\q{\theta}
\def\r{\rho}
\def\s{\sigma}
\def\t{\tau}

\def\L{\Lambda}

\def\S{\Sigma}

\def\dag{\dagger}
\def\pro{\propto}
\def\la{\left}
\def\ra{\right}
\def\pa{\partial}

\def\inf{\infty}

\def\abs#1{\left| #1\right|}

\def\bar#1{\overline{#1}}

\def\ba{\begin{array}}
\def\ea{\end{array}}
\def\be{\begin{equation}}
\def\ee{\end{equation}}
\def\bdm{\begin{displaymath}}
\def\edm{\end{displaymath}}
\def\bea{\begin{eqnarray}}
\def\eea{\end{eqnarray}}
\def\nl{\nonumber \\}

\def\by{\over}
\def\lb{\label}
\def\bl#1{(\ref{#1})}
\def\sp{~~~~~}

\def\ni{\noindent}

\def\pr{Phys.\ Rev.\ }
\def\plb{Phys.\ Lett.\ B }
\def\npb{Nucl.\ Phys.\ B }

\def\cu{\it
Department of Physics, CB 390\\
University of Colorado, Boulder CO 80309
}

\def\br{\begin{thebibliography}{99}}
\def\er{\end{thebibliography}}
\def\rf{\bibitem}
\def\fr#1{\cite{#1}}

\begin{document}
\draft
\preprint{COLO-HEP-353}
\title{Composite Gauge fields and Broken Symmetries}
\author{B. S. Balakrishna and K. T. Mahanthappa}
\address{\cu}
\maketitle
\begin{abstract}
A generalization of the non-Abelian version of the $CP^{N-1}$ models
(also known as Grassmannian models) is presented. The generalization
helps accommodate a partial breaking of the non-Abelian gauge
symmetry. Constituents of the composite gauge fields, in many cases,
are naturally constrained to belong to an anomaly free representation
which in turn generates a composite scalar simulating Higgs mechanism
to break the gauge symmetry dynamically for large $N$. Two cases are
studied in detail: one based on the SU(2) gauge group and the other on
SO(10). Breakings such as SU(2)$\to$U(1) or
SO(10)$\to$SU(5)$\times$U(1) are found feasible. Properties of the
composites fields and gauge boson masses are computed by doing a
derivative expansion of the large $N$ effective action.
\\
\\
\ni PACS nos: 11.15.Ex, 12.50.Fk
\end{abstract}

\section{Introduction}
\lb{int}

Compositeness has been the way of nature. Some of the so called
elementary particles of earlier times have turned out to be composites
of more elementary ones. Compositeness is one way of getting to a
simpler theory involving usually fewer fields or fewer parameters at
the fundamental level. It is also expected to soften the ultraviolet
behavior. Composite gauge fields could in addition provide us with an
understanding of the gauge principle. They have received considerable
attention in the literature\fr{old,cpn,gms,pal}. The models that are
relevant to our purpose are the so called $CP^{N-1}$ models\fr{cpn}
and their non-Abelian generalizations called Grassmannian
models(GMs)\fr{gms}. Here the gauge fields arise as composites of
bosonic constituents\fr{pal}. These models have been studied in the
large $N$ limit and it is found that the non-Abelian symmetry is
either completely broken or not broken at all. Ultimately one would
like to construct phenomenological models along these lines, but the
phase structure of these models is not very useful for that purpose.
One needs a version in which the gauge symmetry is partially broken.

Remarkably, as we will see in this article, there does exist a
generalization of the GMs that allows for partial symmetry breaking.
Some of the results of this paper have been briefly reported
earlier\fr{our}.  The primary agent of symmetry breaking turns out to
be a scalar that too is composite. This composite Higgs scalar arises
naturally as a solution of the modified constraint equation. In many
of the cases, it belongs to the adjoint representation of the gauge
group. The constituent fields in those cases belong to an anomaly free
representation. One may recall here that that the agent of symmetry
breaking in grand unified theories is usually a Higgs scalar in the
adjoint representation. One may further recall that the fermions in a
physical theory belong to an anomaly free representation, and we
encounter the same feature here, though in the bosonic version. Two
examples are studied to illustrate the approach: one based on the
gauge group SU(2) and the other based on SO(10). The SU(2) example is
the simplest and best suited to illustrate the approach. Here there
exists a phase where SU(2) breaks to a U(1) subgroup. The case of
SO(10) studied in some detail is more interesting from the physical
point of view. The phase structure is richer with symmetry breaking to
various subgroups such as SU(5) or SU(5)$\times$U(1).

We also compute the properties of the composite fields, the gauge
bosons and the Higgs scalars, by doing a derivative expansion of the
large $N$ effective action. The expansions available in the literature
do not serve our purpose as they are, to our knowledge, also
expansions in the Higgs scalar. We hence develop a suitable derivative
expansion which we use to compute the kinetic terms and the mass terms
for the composites in the various phases.

To start with, in the next section, we review the known models
relevant to our work. First we look at the $CP^{N-1}$ model that
involves a U(1) gauge theory. Then we discuss its non-Abelian
generalization, the Grassmannian model. It is based on the gauge group
U($M$) with the scalars in the fundamental representation. In section
\ref{gen}, we introduce our model, a generalization of GM that is
capable of accommodating other gauge groups with more general scalar
representations. A suitable potential responsible for a rich phase
structure is then introduced. In section \ref{two}, we first
illustrate our approach with a simple example based on the gauge group
SU(2) and then look at the interesting but more complicated case of
SO(10). Section \ref{pro} discusses the properties of the various
composites. The global symmetry with its breaking patterns and the
resulting Goldstone modes are discussed in section \ref{glo}. Section
\ref{dis} concludes with a discussion of the present approach.
Derivative expansion of the effective potential useful in section
\ref{pro} is carried out in appendix \ref{der}.

\section{Known models}
\lb{kno}

The known models that are relevant to our purpose, the $CP^{N-1}$ and
the Grassmannian models, are briefly reviewed in this section. The
simplest model is the one that induces a U(1) gauge theory, the
$CP^{N-1}$ model. This is just a field theory involving $N>1$ complex
scalar fields,
\be
Z = \la(Z_1,Z_2,\cdots,Z_N\ra),
\ee
satisfying the constraint $\sum_i\abs{Z_i}^2=1$.  The constraint sets
their overall scale. A U(1) gauge invariance removes in addition an
angular variable. Thus the model is in effect a field theory of $N-1$
complex scalars. For convenience in writing the various results, we
have above represented the $Z_i$'s collectively as a row vector $Z$.
In this notation, the constraint can be rewritten as $ZZ^\dag=1$. The
Lagrangian is
\be
L = \b N\la[\pa_\m Z\pa_\m Z^\dag + \la(Z\pa_\m Z^\dag\ra)^2\ra].
\lb{l1}
\ee
Here $\b$ is the inverse of a coupling constant. An overall
multiplicative factor $N$ is introduced for later convenience in the
$1/N$ expansion.  It is easy to verify the existence of a $U(1)$ gauge
invariance under which each of the $Z_i$'s transform with the same
phase. In the vector notation, this is simply $Z\to e^{i\q}Z$ where
the phase $\q$ is space-time dependent. The constraint is clearly
invariant under this symmetry. To see that it is a gauge symmetry,
note first that the combination $iZ\pa_\m Z^\dag$ transforms as a U(1)
gauge field,
\be
iZ\pa_\m Z^\dag \to iZ\pa_\m Z^\dag + \pa_\m\q.
\ee
The last term actually has a $ZZ^\dag$ but that drops out due to the
constraint. The U(1) gauge invariance will be more explicit, if we
rewrite the Lagrangian by introducing an auxiliary field
$A_\m=iZ\pa_\m Z^\dag$ as follows:
\be
L = \b N\la(\pa_\m Z\pa_\m Z^\dag - 2iA_\m Z\pa_\m Z^\dag + A_\m^2\ra)
= \b N\la[D_\m Z(D_\m Z)^\dag\ra].
\lb{l2}
\ee
Here $D_\m Z$ is the covariant derivative $(\pa_\m-iA_\m)Z$. In this
form, the gauge symmetry is manifest. As shown in section \ref{pro},
the auxiliary field $A_\m$ that transforms as a $U(1)$ gauge field
becomes dynamical and hence a genuine gauge field after quantum
corrections in the large $N$ approximation. It is a composite gauge
field made of the $Z$ fields.  Thus the model under consideration can
be viewed as an induced $U(1)$ gauge theory or a theory of composite
gauge fields. This model is a special case of the Grassmannian model;
hence we study its phase structure below as a special case of GM.

The Grassmannian model is a generalization of the $CP^{N-1}$ model
that induces a non-Abelian gauge theory. We now have more fields, a
set of them, represented collectively by a $M\times N$ matrix $Z$ with
the elements $Z_{\a i}$, $\a$ labeling the rows and $i$ labeling the
columns. The column index $i$ is an internal index or a flavor index
that is essentially carried over from our previous model. The new
index $\a$, the row index, is the gauge index associated with a
non-Abelian symmetry which in the present case is U($M$). All our
results should reduce to those of the previous model for the case of
$M=1$. The constraint is now
\be
ZZ^\dag = I_M,
\ee
where $I_M$ is an identity matrix of order $M$. The present Lagrangian
is of the previous form \bl{l1}, but now as such it will be an
$M\times M$ matrix and hence needs an overall trace to make it a
number,
\be
L = \b N{\rm tr}\la[\pa_\m Z\pa_\m Z^\dag + \la(Z\pa_\m
Z^\dag\ra)^2\ra].
\ee
It is again easy to verify that there is a U$(M)$ gauge invariance
with respect to the index $\a$. Under this gauge symmetry, $Z$
transforms as a set of $N$ fundamental representations. In the matrix
notation, this transformation is simply $Z\to UZ$, $U$ being an
$M\times M$ space-time dependent unitary matrix representing the gauge
transformation. This transformation does not affect the $i$ index
which labels $N$ fundamental representations. The constraint respects
this symmetry. The object $A_\m=iZ\pa_\m Z^\dag$ is in the adjoint
representation of U($M)$ and transforms as a gauge field thanks to the
constraint,
\be
A_\m \to UA_\m U^\dag + iU\pa_\m U^\dag.
\lb{gtm}
\ee
The role of the constraint here is to simplify the last term above
from $iUZZ^\dag\pa_\m U^\dag$ to $iU\pa_\m U^\dag$. The gauge symmetry
becomes explicit when we rewrite the Lagrangian as in Eq. \bl{l2} with
an overall trace,
\be
L = \b N{\rm tr}\la[D_\m Z(D_\m Z)^\dag\ra].
\lb{l3}
\ee
$D_\m Z$ being the covariant derivative $(\pa_\m-iA_\m)Z$. As before,
$A_\m$ appears as an auxiliary field but, as can be seen at large $N$,
it becomes dynamical and hence a genuine gauge field after quantum
corrections. It is a composite gauge field with the $Z$ fields as
constituents. The constraint and the $U(M)$ gauge invariance have the
effect of suppressing $M^2$ degrees of freedom. The theory is thus
based effectively on $M(N-M)$ scalars.  Clearly, for it to be a
sensible one, $N$ is required to exceed $M$.

The constraint $ZZ^\dag=I_M$ can be incorporated into the Lagrangian
with the help of a Lagrangian multiplier $\S$, a $M\times M$ matrix.
The result is
\be
L = \b N{\rm tr}\la[D_\m Z(D_\m Z)^\dag + \S ZZ^\dag - \S\ra].
\lb{l4}
\ee
To understand symmetry breaking, and hence to identify the various
phases, we need to obtain the effective potential. We will do this at
large $N$. Because $A_\m$ is not expected to pick up any expectation
value, we will set it to zero. The classical contribution to the
effective potential comes from the Lagrangian \bl{l4} by dropping the
derivative terms. Because $1/N$ appears in the Lagrangian like the
Planck's constant, the quantum corrections to this contribution is
expected to be suppressed by a factor $1/N$. But there are $N$
fundamental representations contributing equally and this can offset
the $1/N$ suppression. The result is that at large $N$ the effective
potential for the $Z$ and $\S$ fields obtained by integrating away the
$Z$ fluctuations carries a correction
\be
N\int{d^4k\by(2\pi)^4}{\rm tr}~{\rm ln}\la(k^2I_M+\S\ra).
\ee
Here and in the rest of this article, we suppress the dependence of
the momentum integrals on a cutoff $\L$. The total effective potential
is thus
\be
V_{\rm eff} = \b N{\rm tr}\la(\S ZZ^\dag-\S\ra) +
N\int{d^4k\by(2\pi)^4}{\rm tr}~{\rm ln}\la(k^2I_M+\S\ra).
\lb{veff}
\ee

To determine the various phases, we need to extremize this potential.
The resulting saddle point equations (SPEs) are $\S Z=0$ obtained by
varying $Z^\dag$ and
\be
\b\la(ZZ^\dag -I_M\ra) + \int{d^4k\by(2\pi)^4}{1\by k^2I_M+\S} = 0
\lb{seq}
\ee
coming from varying $\S$. Now, let us look for solutions of the form
\be
\S = \la(\ba{cc}\s.I_p&0\\0&0\ea\ra), \sp ZZ^\dag =
\la(\ba{cc}0&0\\0&v^2.I_{M-p}\ea\ra),
\ee
where $I_p$ and $I_{M-p}$ are two identity matrices of order $p$ and
$M-p$ respectively. Note that $\S ZZ^\dag$ is zero with this ansatz.
Solutions for $Z$ that satisfy $\S Z=0$ can easily be constructed. Eq.
\bl{seq} leads to two equations, one in the $I_p$ sector and the other
in the $I_{M-p}$ sector,
\bea
-\b + \int{d^4k\by(2\pi)^4}{1\by k^2+\s} &=& 0, \nl
\b(v^2-1) + \int{d^4k\by(2\pi)^4}{1\by k^2} &=& 0.
\lb{sp}
\eea
For $p=0$ there is no $I_p$ sector and hence the first equation above
would be absent. Similarly for $p=M$ the second would be absent. Let
us first show that $p$ can not lie in between. We will do this by
showing that the two equations can not be satisfied simultaneously.
Firt note that $\s$ should not be negative for the momentum integral
involving it to be well-defined.  Hence, from the first equation, we
note that $\b$ has an upper limit $\b_c$ given by
\be
\b_c = \int{d^4k\by(2\pi)^4}{1\by k^2} = {\L^2\by 16\pi^2}.
\lb{bc}
\ee
But the second equation implies that $\b_c$ is also a lower limit of
$\b$. To see this, rewrite it as $v^2=1-\b_c/\b$ and note that $v^2$
can not be negative. It thus follows that, except at the critical
point $\b=\b_c$ where the question is irrelevant, $p$ is either zero
or $M$.

Thus we have two phases. For $\b>\b_c$, we have the broken phase
($p=0$) where $ZZ^\dag$ has an expectation value. The solution for $Z$
has an expectation value along all the ``directions'' in the
fundamental representation. This breaks $U(M)$ completely and all the
gauge bosons are massive. For $\b<\b_c$, we have the unbroken phase
($p=M$) where the gauge symmetry is unbroken and the gauge bosons are
massless. $\b=\b_c$ is a critical point. In fact, $\b\leq\b_c$ is a
critical line along which all the masses vanish.

In other words, the gauge group is either completely broken or not
broken at all. There are apparently no phases, at least at large $N$,
where a partial breaking of the gauge group is possible. To obtain a
richer phase structure, we invoke a generalization of these models and
study them at large $N$ in the following section.

\section{A Generalization}
\lb{gen}

As discussed in the previous section, the Grassmannian model involves
scalars in $N$ fundamental representations of the gauge group U($M$).
A natural extension is to construct models for various gauge groups
with the scalars transforming under different representations.  To our
knowledge, they have not been studied in the literature. In this
section, we explore an interesting class of these models that are
nontrivial generalizations offering a rich phase structure. As usual,
we are concerned with symmetry breaking at large $N$.

\subsection{Modifying the constraint}

One approach, a straightforward one, is to choose some other gauge
group $G$ in place of U($M$) but to leave the constraint $ZZ^\dag=I_M$
unchanged. In other words, we take the $Z$ fields to belong to an
arbitrary representation $R$ of dimension $M$ and multiplicity $N$ of
some chosen gauge group $G$. We may still represent the $Z$ fields in
the form of an $M\times N$ matrix. The transformation matrix $U$ is
now in $R$ acting on the matrix $Z$ as before, $Z\to UZ$. The
Lagrangian is still of the form we came across earlier in Eq. \bl{l3},
but with the auxiliary gauge field $A_\m$ now taking values in the Lie
algebra of $G$. An expression for the auxiliary field and the form of
the Lagrangian involving the $Z$ fields alone can be easily derived.
They are given respectively by the Eqs. \bl{amu} and \bl{l5} given
below. However, symmetry breaking at large $N$ remains the same. This
is because the gauge fields are set to zero in our discussion of the
phase structure.

A more interesting generalization occurs when the constraint is
modified as well. Again, we take the $Z$ fields to be in any
representation $R$ of dimension $M$ and multiplicity $N$ of a gauge
group $G$. We look for a Lagrangian that resembles \bl{l3}. It is
clearly gauge invariant with the auxiliary gauge field $A_\m$
transforming as in \bl{gtm}. Note that the part of the Lagrangian
quadratic in $A_\m=A_\m^aT_a$ is proportional to
\bdm
A_\m^aA_\m^b{\rm tr}(T_aT_bZZ^\dag)=A_\m^aA_\m^b{\rm
tr}(T_{ab}ZZ^\dag),
\edm
where $T_a$'s are the generators of the gauge group $G$ and
$T_{ab}=(T_aT_b+T_bT_a)/2$. Earlier in the GM, the constraint
$ZZ^\dag=I_M$ was responsible for rendering it quadratic in $A_\m$
alone. This resulted in a well defined expression for the auxiliary
field $A_\m$ as a composite of the $Z$ fields. Now, more generally, we
achieve the same goal by imposing the following constraint instead:
\be
{\rm tr}(T_{ab}ZZ^\dag) = l\d_{ab}.
\lb{con}
\ee
Here $l$ is the Dynkin index of the representation $R$ defined by
${\rm tr}(T_aT_b)=l\d_{ab}$. Note that this new constraint also
respects the gauge symmetry. Using it, it is easy to obtain the
following expression for the auxiliary field:
\be
A_\m^a = {i\by 2l}{\rm tr}\la[T_a(Z\pa_\m Z^\dag-\pa_\m
ZZ^\dag)\ra].
\lb{amu}
\ee
Later in section \ref{pro}, we observe that this field becomes
dynamical at large $N$ and is justly called a composite gauge field
constructed out of the $Z$'s. An expression for the Lagrangian in
terms of the $Z$ fields alone can now be obtained,
\be
L = \b N{\rm tr}(\pa_\m Z\pa_\m Z^\dag) + {1\by 4l}\b N\la\{{\rm
tr}\la[T_a(Z\pa_\m Z^\dag-\pa_\m ZZ^\dag)\ra]\ra\}^2.
\lb{l5}
\ee
This derivation should ensure gauge invariance and this is easily seen
to be the case. Incorporating the constraint into the Lagrangian using
a Lagrange multiplier $\S=\S_{ab}T_{ab}$ leads to an expression that
agrees with \bl{l4}. The symmetry breaking effects could be
potentially different since the matrix $\S$ is not an arbitrary one
any more.

First, we make sure that the constraint \bl{con} is different from the
earlier one $ZZ^\dag=I_M$. $ZZ^\dag=I_M$ clearly is a solution of
\bl{con}. Fortunately, there are cases where this is not the only
solution. The new constraint is in some cases weaker than the earlier
one. To see this, introduce a hermitian matrix $W$ by $ZZ^\dag=I_M+W$
and observe that the new constraint is equivalent to looking for a
solution of ${\rm tr}(T_{ab}W)=0$. Given a $W$ that leads to a
positive semidefinite $ZZ^\dag$, $Z$ is solvable generally as
$Z=(I_M+W)^{1/2}Z_0$ for some $Z_0$ obeying $Z_0Z_0^\dag=I_M$. The
earlier constraint corresponds to the trivial solution $W=0$. That
there exist cases where $W$ is nontrivial can be seen as follows. Make
the ansatz that $W$ is in the Lie algebra itself, that is $W=W_aT_a$.
Now, the constraint ${\rm tr}(T_{ab}W)=0$ simply states that the
Adler-Bell-Jackiw anomaly associated with the representation $R$
should vanish. That there exist anomaly free representations is well
known. A simple example is the doublet of the gauge group SU(2) that
yields a triplet for $W$. We will use this example later to illustrate
our approach. A more interesting example is the spinor representation
16 of the gauge group SO(10) that is well known to anomaly free. The
ansatz gives a solution for $W$ that is in the adjoint representation
45. This is interesting, for it is known that an adjoint scalar is a
promising candidate to break a grand unified theory based on SO(10).
Its appearance here is quite unexpected.

In some instances, the above ansatz gives the most general solution.
This is the case with both the examples mentioned above. This can be
seen from representation theory. Regard $W$ as belonging to the
product representations $2\times 2=1+3$ and
$16\times\bar{16}=1+45+210$ in the SU(2) and SO(10) examples
respectively. The component representations in these decompositions
should separately obey the constraint. The singlet appearing in both
the cases is ruled out easily. This is because the singlet component
being proportional to identity gives ${\rm tr}(T_{ab}W)\pro \d_{ab}$
violating the constraint. In the case of SO(10), we need to exclude
the 210 as well. It is easily observed that there exist
$16\times\bar{16}$ traceless hermitian matrices violating the
constraint. This excludes 210 because such matrices can only have
components along the 45 and 210, and the 45 alone can not violate the
constraint. It is not to be deduced, however, that the ansatz always
gives the most general solution. For instance, if one were to pick a
sufficiently large representation of the gauge group for $R$, one will
easily end up with more representations that remain unsuppressed in
$W$. But, representation theory should still be applicable to solve
for $W$ in general.

Let us call the models as type one models when $W$ solves identically
to zero and our new constraint reduces to the old one. They are closer
to the Grassmannian models discussed earlier, or rather to their
generalizations mentioned in the beginning of this section involving
subgroups of U($M$). The other models where $W$ can be nontrivial is
referred to as type two models. Models based on a reducible $R$ are
quite generally of type two. This is because the constraint does not
determine some components of $W$, for instance those connecting
different subrepresentations in $R$. Note, however, that a reducible
$R$ arising from an irreducible one repeated say $q$ times, though
appears to be of type two, can be recast as of type one by combining
$q$ and $N$ to an overall multiplicity $qN$ in place of $N$. We prefer
to view them as type one models with multiplicity $qN$. Type one
models are thus necessarily based on irreducible $R$'s.

The effective potential at large $N$ is again of the form \bl{veff}
encountered earlier. The SPEs governing the phases are $\S Z=0$
obtained by varying $Z^\dag$ and
\be
\b{\rm tr}\la[T_{ab}(ZZ^\dag-I_M)\ra] + \int{d^4k\by(2\pi)^4}{\rm
tr}\la(T_{ab}{1\by k^2I_M+\S}\ra) = 0
\ee
obtained by varying $\S$. For type one models, the traces can be
dropped and the equation becomes equivalent to the one we had earlier
for GM (see Eq. \bl{seq}). Solutions can be sought in the same manner.
The phase structure is governed by a critical $\b$ given by Eq.
\bl{bc}. It is ensured that the ansatz for $\S$ one makes in solving
these equations is consistent with its definition $\S=\S_{ab}T_{ab}$.
This is because the final conclusion involves either the broken phase
$\S=0$ (for $\b>\b_c$) or the unbroken phase $\S=\s I_M$ (for
$\b<\b_c$). This is equivalent to $\S_{ab}=0$ or
$\S_{ab}=\s\d_{ab}/C_2(R)$, where $C_2(R)$ is the second Casimir
invariant of the irreducible representation $R$ defined by
$T_aT_a=C_2(R)I_M$, and is acceptable.

For type two models, the above equation can still be reduced to that
of GM, Eq. \bl{seq}, but with a matrix $W$ satisfying ${\rm
tr}(T_{ab}W)=0$ on the r.h.s,
\be
\b\la(ZZ^\dag -I_M\ra) + \int{d^4k\by(2\pi)^4}{1\by k^2I_M+\S} = \b W.
\lb{rsw}
\ee
The factor $\b$ on the r.h.s makes this equation agree with
$ZZ^\dag=I_M+W$ at the level of expectation values. Again, the
solutions $\S=0$ (for $\b>\b_c$) and $\S=\s I_M$ (for $\b<\b_c$) will
satisfy this equation, for $W$ can clearly be chosen zero. As before,
$\S=\s I_M$ is acceptable for irreducible $R$'s as it follows from
$\S_{ab}=\s\d_{ab}/C_2(R)$. In the case of reducible $R$'s, that is
$R=\sum_iR_i$ where each $R_i$ is irreducible, the solution for
$\b<\b_c$ is a bit more involved. Try again the ansatz
$\S_{ab}=\s\d_{ab}$. In this case $\S$ is not proportional to
identity. Instead, it is a diagonal matrix taking values $C_2(R_i)\s$
along each representation $R_i$. We look for a $W$ matrix that is also
diagonal, with values $w_i$ along $R_i$. Note that the constraint
${\rm tr}(T_{ab}W)=0$ requires the $w_i$'s to satisfy $\sum_il_iw_i=0$
where the $l_i$'s are the indices of the representations $R_i$'s. The
SPE along $R_i$ is
\be
-\b + \int{d^4k\by(2\pi)^4}{1\by k^2+C_2(R_i)\s} = \b w_i.
\ee
The constraint on the $w_i$'s gives
\be
-\b\sum_il_i + \sum_il_i\int{d^4k\by(2\pi)^4}{1\by k^2+C_2(R_i)\s} =
0.
\ee
This determines $\s$. Individual equations simply determine the
various $w_i$'s. That this solution holds only for $\b<\b_c$ is easy
to note.

The solutions obtained so far are not a priori the most general ones.
There could be more solutions. This does not, however, appear to be
the case with the examples mentioned earlier based on the gauge groups
SU(2) or SO(10) (see Eqs. \bl{feq2} and \bl{feq10} and the discussions
following them). Analogous situation occurs for the gauge group $E_6$
with its representation 27. This is perhaps illustrative of a generic
phenomenon or suggestive of the need to look at larger representations
that might lead to more solutions. We do not wish to go into those
details here, rather we find it more rewarding to consider the other
possibility, that of adding a potential.

\subsection{Adding a potential}

The unexpected appearance of an adjoint scalar $W$ apparently didn't
help us in a partial breaking of the gauge group. The situation
changes drastically when a potential is introduced leading to a rich
phase structure. The adjoint scalar, that has not played any
significant role so far, plays a major one in the presence of a
potential. Note that there is no simple way to incorporate a potential
in the canonical GM without spoiling the global symmetries or the
constraint equation. But, interestingly, the generalized models of the
previous section, governed by the gauge invariant Lagrangian of Eq.
\bl{l5} constructed with scalars alone, do allow for potential terms.
As we will see, the presence of a potential leads to drastically
different conclusions. These models with such phase structures are
relevant in model building.

Let us keep the potential quite general to begin with, $\b N{\rm
tr}V(ZZ^\dag)$, where $V(\cdot)$ is an ordinary function of its
argument, a polynomial for instance. This is expected to be a
nontrivial extension of type two models unlike the case of type one
models in which the constraint $ZZ^\dag=I_M$ reduces this to the
addition of a constant. It is convenient to introduce a composite
field variable $X$ for $ZZ^\dag$ and write the potential as $\b N{\rm
tr}V(X)$. The requirement $X=ZZ^\dag$ can be incorporated with the
help of a Lagrange multiplier $Y$, adding a term $\b N{\rm
tr}(YZZ^\dag-YX)$ to the potential. As before, constraint \bl{con} can
be accommodated with the help of a Lagrange multiplier
$\S=\S_{ab}T_{ab}$. Its effect is, as we know, to add a term $\b N{\rm
tr}(\S ZZ^\dag-\S)$ to the potential.  After translating $Y$ to $Y-\S$
for convenience, the total Lagrangian looks like
\be
L = \b N{\rm tr}\la[D_\m Z(D_\m Z)^\dag +V(X) +YZZ^\dag -YX +\S X
-\S\ra].
\lb{l6}
\ee
The large $N$ effective potential is now computable,
\be
V_{\rm eff} = \b N{\rm tr}\la[V(X)+YZZ^\dag-YX+\S X-\S\ra] +
N\int{d^4k\by(2\pi)^4}{\rm tr}~{\rm ln}\la(k^2I_M+Y\ra).
\ee
The SPEs are obtained by extremizing this potential.  Varying $X$
determines $Y$ to be $\S+V'(X)$ where a prime denotes differentiation
with respect to the argument. Varying $\S$ gives ${\rm
tr}[T_{ab}(X-I_M)]=0$. As before, one may look for a solution of this
in the form $X=I_M+W$ where $W$ satisfies ${\rm tr}(T_{ab}W)=0$.
Varying $Y$ and using these solutions yields
\be
\b\la(ZZ^\dag -I_M\ra) + \int{d^4k\by(2\pi)^4}{1\by
k^2I_M+\S+V'(I_M+W)} = \b W.
\ee
This is to be supplemented with $YZ=\la[\S+V'(I_M+W)\ra]Z=0$ obtained
by varying $Z^\dag$. This system of equations resembles the ones
obtained earlier (see Eq. \bl{rsw}), with $\S$ replaced by
$\S+V'(I_M+W)$. The presence of $V'(I_M+W)$, however, is suggestive of
a different phase structure.

For type one models $W=0$, and $V'(I_M+W)$ just adds a constant to
$\S$. This can be absorbed into $\S$ because these models, being based
on an irreducible $R$, allow for the addition of term proportional to
identity to $\S$. As expected in the beginning of this section, this
is a trivial extension. However, this is not the case for type two
models and we expect a rich phase structure.

An example we will use later is a potential of sixth degree in $Z$ and
$Z\dag$ that leads to $V'(I_M+W) = aI_M+bW+cW^2$ for some constants
$a$, $b$ and $c$. For a model based on an irreducible $R$, the term
$aI_M$ can be absorbed into $\S$ as we have already noted. When $W$ is
in the adjoint representation, the term $cW^2$ can also be absorbed
into $\S$.  As a result, the SPEs to be solved are
\be
\b\la(ZZ^\dag -I_M\ra) + \int{d^4k\by(2\pi)^4}{1\by
k^2I_M+\S+bW} = \b W,
\ee
and $\la[\S+bW\ra]Z=0$. These equations are difficult to handle
analytically and we present our numerical results for SU(2) and SO(10)
below. We find that they do have solutions for a range of parameters
when $\b<\b_c$ and $b<0$.

Given all such solutions for $\b<\b_c$, the next step is to determine
those that are preferred energetically. The GM solutions of section
\ref{kno} leave the gauge group unbroken for $\b<\b_c$ whereas those
found here break it at least partially. Which one is preferred is of
course determined by the effective potential. In other words, one
needs to compute $V_{\rm eff}$ for all the solutions and pick the one
(or more) that has the the lowest value. We do this along a chosen
path in the parameter space of $\b$ and $b$ that crosses all the
phases. We find that some of the new solutions end up always having
the lowest potential. In other words, for a range of parameters, a
partial breaking of the gauge group is preferred over the unbroken
case.  Details are presented in the next section. The following
expression for the effective potential at a saddle point (SP) is used
to this end:
\be
V_{\rm eff}({\rm SP}) = -\b N{\rm tr}\la(\S+bW^2/2\ra) +
N\int{d^4k\by(2\pi)^4}{\rm tr}~{\rm ln}\la(k^2I_M+\S+bW\ra).
\ee
We are, as always, concerned with an adjoint $W$ for an irreducible
$R$. In the above expression, terms $aI$ and $cW^2$ have been absorbed
into $\S$ for convenience.

\section{Two examples: SU(2) and SO(10)}
\lb{two}
\subsection{The case of SU(2)}

\begin{figure}
\vspace {0.1\textheight}
\epsfxsize=0.65\textwidth
\epsffile{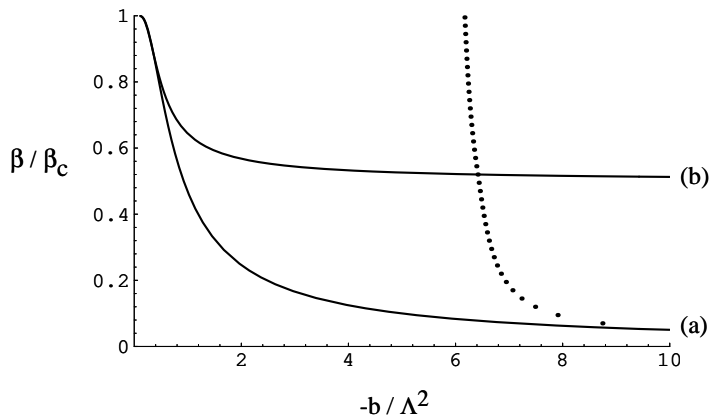}
\caption{The phase diagram in the case of SU(2) obtained solving Eqs.
(29) and (31). The effective potential of Fig. 2 is computed along the
dotted line. Parameters in the theory are $\b$ and $b$, and $\L$ is
the momentum cutoff. Details are given in the text.}
\label{pha1}
\end{figure}

The example based on SU(2) with doublet $Z$'s is the simplest and the
most convenient one to illustrate the ideas presented above. The
matrix $\S=\S_{ab}T_{ab}$ is now proportional to identity, hence
chosen to be $\s I_2$. The $W$ scalar, being a triplet, is taken to be
along the $\s_3$ direction, that is $W=w\s_3$. First we consider the
case when the $Z$ fields develop no expectation value. The resulting
SPEs in the presence of a potential are
\bea
-\b + \int{d^4k\by(2\pi)^4}{1\by k^2+\s+bw} &=& \b w, \nl
-\b + \int{d^4k\by(2\pi)^4}{1\by k^2+\s-bw} &=& -\b w.
\lb{feq2}
\eea
Note that there is no solution to these equations when $b=0$, that is
in the absence of a potential, other than the one discussed earlier
where $W=0$ and $\S$ is proportional to identity. In the presence of a
sixth degree potential given by $V'(I_M+W) = aI_M+bW+cW^2$, these
equations do have solutions for a range of parameters. This can be
seen by treating $x=(\s+bw)/\L^2$ and $y=(\s-bw)/\L^2$ as independent
variables to determine $\b$ and $w$ from the above two equations.
Given $x$ and $y$ and knowing $w$, one obtains $b$ from
$x-y=2bw/\L^2$. The resulting equations are
\bea
1-{\b\by\b_c} &=& {1\by 2}x~{\rm ln}\la(1+{1\by x}\ra) + {1\by
2}y~{\rm ln}\la(1+{1\by y}\ra) \nl
-{\b_c\by\b}{b\by\L^2} &=& {x-y\by x~{\rm ln}\la(1+1/x\ra) - y~{\rm
ln}\la(1+1/y\ra)}.
\lb{feq}
\eea
For the momentum integrals to remain well defined, $x$ and $y$ should
be positive (or zero). The region of the parameter space of $\b$ and
$b$ is obtained by letting $x$ and $y$ vary from zero to infinity.
This falls in between the curves (a) and (b) shown in Fig.
\ref{pha1}. There are two solutions for a given $\b$ and $b$ in this
regime, but they are related to each other by an interchange of $x$
and $y$ and should be treated as one. Note that all these solutions
yield $\b<\b_c$ and $b<0$. Curve (a) has $x=y$ and is the critical
line. In fact, the region below curve (a) is a critical surface where
all the masses vanish. Curve (b) has one of $x,y$ zero.  Symmetry
breaking involved here is from SU(2) to U(1).

\noindent

There exist new solutions for a nonzero $Z$ as well. Giving an
expectation value diag($0,v^2$) for $ZZ^\dag$, the second equation in
\bl{feq2} gets replaced by
\be
\b v^2 -\b + \b_c = -\b w,
\lb{leq2}
\ee
where we have set $y=\s-bw$ to zero to satisfy $(\S+bW)Z=0$. Treating
$x=(\s+bw=2bw)/\L^2$ and $y'=v^2\b/\b_c$ as independent variables to
determine the others,
\bea
1-{\b\by\b_c} &=& {1\by 2}x~{\rm ln}\la(1+{1\by x}\ra) -{1\by 2}y' \nl
-{\b_c\by\b}{b\by\L^2} &=& {x\by x~{\rm ln}\la(1+1/x\ra) + y'},
\eea
one notes the presence of solutions in a parameter range for positive
$x$ and $y'$. Here as well, we require $\b<\b_c$ and a negative $b$.
The parameter range is the one above curve (b) in Fig. \ref{pha1}.
SU(2) symmetry is now completely broken. Giving an expectation value
diag($v^2,0$) for $ZZ^\dag$ is equivalent to this case and leads to no
new solutions. A nonzero $ZZ^\dag$ of the form diag($v_1^2,v_2^2$)
requires $v_1^2=v_2^2$ and coincides with the completely broken case
of GM discussed in section \ref{kno}.

\begin{figure}
\vspace {0.1\textheight}
\epsfxsize=0.65\textwidth
\epsffile{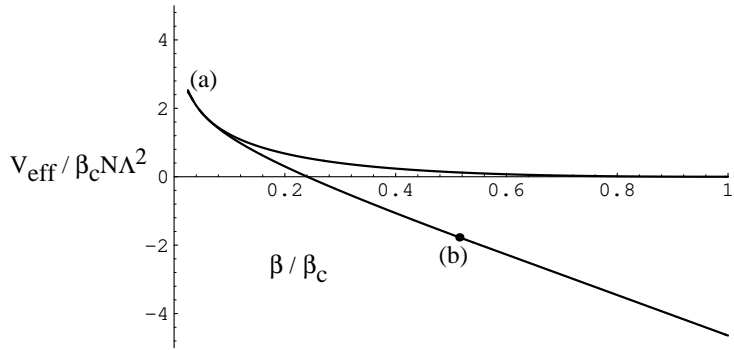}
\caption{A plot of the effective potential (with its zero
appropriately chosen) versus $\b$ for a path shown dotted in Fig. 1
crossing all the curves. The crossings are denoted by (a) and (b). The
upper curve corresponds to the unbroken case and the lower one
corresponds to symmetry breaking as discussed in the text.}
\label{pot1}
\end{figure}

These are all the solutions. Now, consider all those for $\b<\b_c$.
The corresponding one of GM leaves the gauge group unbroken whereas
those of this section break it at least partially. As discussed
earlier, the effective potential needs to be examined to determine the
preferred solution. We have chosen a path suitably fixing $y$ crossing
all the curves, shown dotted in Fig. \ref{pha1}. Fig. \ref{pot1} is a
plot of the effective potential. The upper curve is for the GM
solutions and the lower one is for the new solutions.  Note that the
lower sheet ends up always having the lowest potential.  In other
words, for $\b$ small, a partial breaking of the gauge group is
preferred over the unbroken case.

\noindent

\subsection{The case of SO(10)}

We now come to the second example, the gauge group SO(10) with the
$Z$'s in the representation 16. Here the possibilities for symmetry
breaking are too many. We do not hope to address all of them, rather
simply pick one possibility: SO(10) breaking to SU(5) or to its
maximal subgroup SU(5)$\times$ U(1). Under SU(5), 16 of SO(10)
decomposes to $10(1)+\bar{5}(-3)+1(5)$ where the U(1) charges are
given in paranthesis. The unbroken symmetry for $Z=0$ corresponds to
those generators that commute with the ansatz for $\S$. To have
symmetry breaking to SU(5)$\times$U(1), we hence choose $Z=0$ and take
$\S_{ab}=\s\d_{ab}$ along the SU(5) directions, $\r$ along the U(1),
zero otherwise. Note that $a$ or $b$ index runs over the adjoint
representation 45 of SO(10) that under SU(5) decomposes to
$24+10+\bar{10}+1$. Our ansatz for $\S_{ab}$ corresponds to having it
nonzero for $(a,b)$ along $(24,24)$ and $(1,1)$. One could have it
nonzero along $(10,\bar{10})$ and $(\bar{10},10)$ as well, but it
turns out that this can be absorbed into $\s$ and $\r$. This means
that the $\S$ matrix is diagonal with values $C_2(10)\s+\r$,
$C_2(\bar{5})\s+9\r$ and $25\r$ along the representations 10,
$\bar{5}$ and 1 respectively. With $C_2(10)/C_2(\bar{5})=3/2$ and a
suitable scaling of $\s$, we may take them to $3\s+\r$, $2\s+9\r$ and
$25\r$. The $W$ matrix is taken to be along the U(1) direction; in
other words, it is diagonal with values $w$, $-3w$ and $5w$. It is now
straightforward to write down the SPEs in the presence of a potential,
\bea
-\b + \int{d^4k\by(2\pi)^4}{1\by k^2+3\s+\r+bw} &=& \b w, \nl -\b +
\int{d^4k\by(2\pi)^4}{1\by k^2+2\s+9\r-3bw} &=& -3\b w, \nl -\b +
\int{d^4k\by(2\pi)^4}{1\by k^2+25\r+5bw} &=& 5\b w.
\lb{feq10}
\eea
Consider first the case of no potential, that is $b=0$. There are no
new solutions. This is because the $\S$ eigenvalues $3\s+\r$,
$2\s+9\r$ and $25\r$ are either in the ascending order or in the
descending order (or equal) and hence one cannot obtain the
alternating signs on the r.h.s above. Allowing for nonvanishing $Z$
does not improve the situation. Note that $ZZ^\dag$ and $\S$ should
have nonnegative eigenvalues and $\S ZZ^\dag=0$ requires at least
$3\s+\r$ or $25\r$ to vanish to allow for a nonzero eigenvalue of
$ZZ^\dag$.

\begin{figure}
\vspace {0.1\textheight}
\epsfxsize=0.65\textwidth
\epsffile{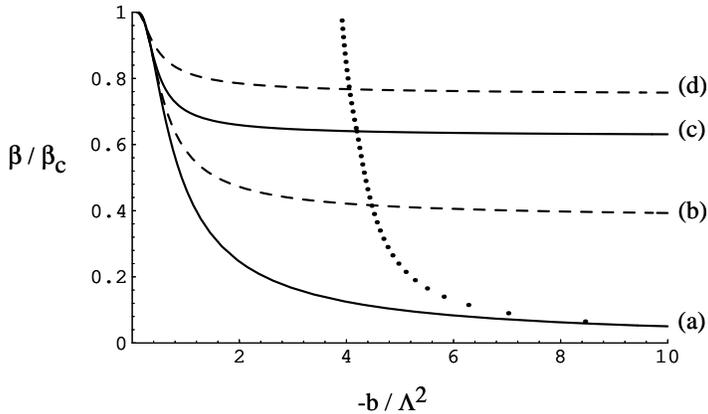}
\caption{The phase diagram in the case of SO(10) obtained solving Eqs.
(33) and (36). The effective potential of Fig. 4 is computed along the
dotted line. Parameters in the theory are $\b$ and $b$, and $\L$ is
the momentum cutoff. Details are given in the text.}
\label{pha2}
\end{figure}

Solutions exist in the presence of a potential for a range of
parameters. To see this, as in the SU(2) case, treat
$x=(3\s+\r+bw)/\L^2$ and $y=(2\s+9\r-3bw)/\L^2$ as independent
variables to determine $\b$ and $w$ from the first two equations and
$z=(25\r+5bw)/\L^2$ from the last one. Given $x$ and $y$ and knowing
$z$ and $w$, one obtains $b$ from $2x-3y+z=16bw/\L^2$. Thus Eq.
\bl{feq10} yields
\bea
1-{\b\by\b_c} &=& {3\by 4}x~{\rm ln}\la(1+{1\by x}\ra) + {1\by
4}y~{\rm ln}\la(1+{1\by y}\ra) \nl
-{\b_c\by\b}{b\by\L^2} &=& {(2x-3y+z)/4\by x~{\rm ln}\la(1+1/x\ra) -
y~{\rm ln}\la(1+1/y\ra)},
\eea
where $z$ is a solution of
\be
z~{\rm ln}\la(1+{1\by z}\ra) = 2x~{\rm ln}\la(1+{1\by x}\ra) - y~{\rm
ln}\la(1+{1\by y}\ra).
\ee
Note that $x$, $y$ and $z$ are required to remain positive (or zero)
to keep the momentum integrals well defined. Solutions exist in a
certain domain of $x$ and $y$ giving rise to a range for the
parameters for $\b<\b_c$ and $b<0$\fr{hig}. The results of our
numerical investigation is presented in Fig. \ref{pha2}. There are in
fact two solutions for a given $\b$ and $b$ in the region between the
curves (a) and (b), and one solution between the curves (b) and (c).
In other words, one of the solutions extends from curve (a) to curve
(b) while the other from curve (a) to curve (c). Curve (a) has $x=y=z$
and is the critical line. Here too, the region below curve (a) is a
critical surface wherein all the masses vanish. Curve (b) has $z=0$
and curve (c) has $y=0$. The symmetry breaking involved here is from
SO(10) to SU(5)$\times$U(1) as noted before.

\noindent

Solutions exist for nonzero $Z$. Consider giving an expectation value
$v^2$ for $ZZ^\dag$ along the singlet in the decomposition
$16=10+\bar{5}+1$. In this case, the third equation in \bl{feq10} gets
replaced by
\be
\b v^2 -\b + \b_c = 5\b w,
\lb{leq10}
\ee
where we have set $z=(25\r+5bw)/\L^2$ to zero to satisfy $(\S+bW)Z=0$.
Again, treat $x$ and $y$ as independent variables and determine the
others to obtain new solutions for positive $v^2$ in a range of
parameters:
\bea
1-{\b\by\b_c} &=& {3\by 4}x~{\rm ln}\la(1+{1\by x}\ra) + {1\by
4}y~{\rm ln}\la(1+{1\by y}\ra) \nl
-{\b_c\by\b}{b\by\L^2} &=& {(2x-3y)/4\by x~{\rm ln}\la(1+1/x\ra) -
y~{\rm ln}\la(1+1/y\ra)},
\eea
where we require
\be
{\b\by\b_c}v^2 = -2x~{\rm ln}\la(1+{1\by x}\ra) + y~{\rm
ln}\la(1+{1\by y}\ra) \geq 0.
\ee
Here as well, one has $\b<\b_c$ and $b<0$. The region of the parameter
space covered by these solutions (one solution for a given $\b$ and
$b$) is that in between the dashed curves (b) and (d) of Fig.
\ref{pha2}. Curve (d) has $x=0$. The surviving symmetry here is SU(5)
because a nonvanishing $Z$ along the singlet in the decomposition
$16=10+\bar{5}+1$ breaks the U(1) subgroup of SU(5)$\times$U(1) as
well. There are other possibilities.  Giving an expectation value for
$ZZ^\dag$ along $\bar{5}$ (instead of the singlet) also leads to a
solution which falls above the curve (c); solutions are also noted to
exist for a nonzero $ZZ^\dag$ along 10 and 1 (extending above curve
(d)), or 10 and $\bar{5}$ (extending beyond that of $\bar{5}$). All of
these, however, break the gauge group completely.

What we have in the end is a two sheeted cover of the parameter space
above the critical curve (a) in Fig. \ref{pha2}. One of them (call it
the upper sheet) is through the solid curves while the other one (call
it the lower sheet) is through the dashed curves. They meet along
curve (a).  There is of course one more sheet (call it the top sheet)
for the solutions of our earlier case of the unbroken gauge group
covering all of the parameter space for $\b<\b_c$. This too meets the
other two sheets along curve (a). For every point on any one of the
sheets, there is a solution.

As we have noted earlier, there could be more solutions. For instance,
there is the possibility that a solution breaking SO(10) to
SU(3)$\times$SU(2)$\times$U(1) (perhaps, with an additional U(1))
exists. The number of variables and the number of equations at least
match, each being six, but the number of equations makes analysis
complicated. There are more possibilities such as SO(10)$\to$SU(4),
SU(4)$\times$U(1), etc. Each case has to be handled separately; a
general treatment has eluded us.

\begin{figure}
\vspace {0.1\textheight}
\epsfxsize=0.65\textwidth
\epsffile{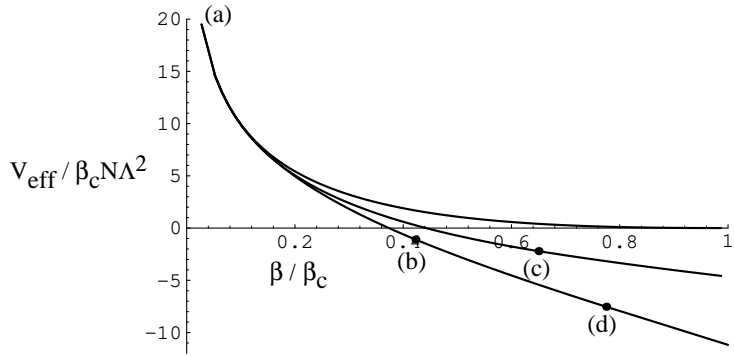}
\caption{A plot of the effective potential (with its zero
appropriately chosen) versus $\b$ for a path shown dotted in Fig. 3
crossing all the curves. The crossings are denoted by (a), (b), (c)
and (d). The uppermost curve corresponds to the unbroken case and the
lower two correspond to symmetry breaking as discussed in the text.}
\label{pot2}
\end{figure}

As before, which solution is preferred is determined by the effective
potential. For this, one needs to compute $V_{\rm eff}$ for all the
solutions and pick the one (or more) that has the the lowest value.
In the present case, this is not an easy task given the number of
possibilities involved. Hence we will be content with doing this
numerically for the solutions found above. We have chosen a path
suitably fixing $y$ in the lower sheet and $z$ in the upper sheet
crossing all the curves, shown dotted in Fig. \ref{pha2}. Fig.
\ref{pot2} is a plot of this effective potential for the three sheets
involved. The uppermost curve is for the top sheet, the middle one is
for the upper sheet and the lowermost one is for the lower sheet. Note
that the lower sheet ends up always having the lowest potential. In
other words, for $\b<\b_c$ but not close to it, a partial breaking of
the gauge group is preferred over the unbroken case.

\noindent

\section{Properties of the composites}
\lb{pro}

Here we study the various composites encountered earlier, the gauge
bosons and composite Higgs particles. Their properties can be read off
from the effective action. Like the effective potential, the effective
action at large $N$ is obtained by integrating away the $Z$ fields. It
involves the original action plus a correction of the form
\be
S_{eff} = -N{\rm ~Tr~ln}\la(-D^2+\S\ra),
\ee
where `Tr' stands for a complete trace, that is, a trace over the
internal indices and an integral over the space-time coordinates. $D$
is as before the covariant derivative $D_\m=\pa_\m-iA_\m$. For
simplicity, we have absorbed the $bW$ term into $\S$. To identify the
kinetic terms for the various fields, a derivative expansion is
needed. This is carried out in Appendix \ref{der}. The result is that
the kinetic terms for the $\S$ fields arise from
\be
{N\by 2}\int d^4x\int {d^4k\by(2\pi)^4}\int_0^{\inf}{dt\by
t}e^{-tk^2}\int_0^tdt_1{\rm tr~}{\cal D}_\m e^{-(t-t_1)\S}{\cal D}_\m
e^{-t_1\S}.
\lb{kts}
\ee
where ${\cal D}_\m{\cal O}$ is $[D_\m,{\cal O}]$ for any ${\cal O}$.
At the next order in the expansion, one obtains an analogous result
for the gauge fields:
\be
{N\by 2}\int d^4x\int{d^4k\by(2\pi)^4}\int_0^{\inf}{dt\by
t^2}e^{-tk^2}\int_0^tdt_1t_1(t-t_1){\rm
tr~}F_{\m\n}e^{-(t-t_1)\S}F_{\m\n}e^{-t_1\S}.
\lb{ktg}
\ee
These general results are valid for any gauge group.

The two phases of GM, the completely broken and unbroken ones
discussed in section \ref{kno}, can be handled in this generic setup.
In the unbroken phase where $\S=\s I,Z=0$ the kinetic terms for the
gauge fields simplifies to
\be
{1\by 2g^2(\s/\L^2)}\int d^4x~{\rm tr~}F_{\m\n}^2
\ee
where $g^2$ is the coupling constant:
\bea
{1\by g^2(\s/\L^2)} &=& {N\by 6}\int{d^4k\by(2\pi)^4}{1\by(k^2+\s)^2}
\nl
&=& {N\by 96\pi^2}\la[{\rm ln}\la(1+\L^2/\s\ra)-{1\by 1+\s/\L^2}\ra].
\lb{fg2}
\eea
This also follows from the well-known contribution to the running
gauge coupling constant at one loop order in the presence of $N$
fundamental scalars. In the broken phase where
$\S=0,ZZ^\dag=(1-\b_c/\b)I_M$, the above computation of the induced
kinetic terms for the gauge fields suffers from an infrared
divergence. Introducing an infrared cutoff $\m$ for $k^2/\L^2$, we
obtain a result that coincides with the above one with $\s/\L^2$
replaced by $\m$. The mass terms arise from the kinetic terms for the
$Z$ fields that reads
\be
\b N{\rm tr}\la(A_\m^2ZZ^\dag\ra) = N(\b-\b_c){\rm tr}\la(A_\m^2\ra).
\ee
The mass squared for $A_\m$ is then $N(\b-\b_c)g^2(\m)$. As
expected, it is positive for $\b>\b_c$ and vanishes at the critical
point $\b=\b_c$.

We now turn our attention to type two models, in particular to the
models based on SU(2) and SO(10). The mass squared results we obtain
are all expressible in units of $\L^2$. All our results for the
kinetic terms and mass terms involve the expectation values of the
scalar field $\S$ through $x$ and $y$ (and in the case of SO(10) $z$
as well) parameters. In the phase where $ZZ^\dag\neq 0$, the
expectation value of $ZZ^\dag$ also appears in the expressions through
$v^2$. As we have seen in the previous section, all these parameters
are expressible in terms the basic ones, $\b/\b_c$ and $b/\L^2$ by
solving a set of equations. When solving the equations however, we
found it convenient to treat $x$ and $y$ as independent variables to
determine $\b/\b_c$ and $b/\L^2$. This helped us to discover multiple
solutions; but once we have chosen the energetically preferred
solution, the relation between the $x,y$ parameters and the basic ones
is one to one and is hence invertible.

\subsection{The case of SU(2)}

First we consider SU(2) and compute the induced kinetic terms for the
gauge fields in the U(1) phase where $\S=\s I_2+bw\s_3$ and $Z=0$.
Writing
\be
F_{\m\n} = {1\by\sqrt{2}}F_{\m\n}^+\s_+ + {1\by\sqrt{2}}F_{\m\n}^-\s_-
+ {1\by 2}F_{\m\n}^3\s_3,
\ee
where $\s_{\pm}=(\s_1\pm i\s_2)/2$, we obtain the kinetic terms for
the gauge bosons of the unbroken U(1) and for those of the broken
generators from Eq. \bl{ktg}. We have
\be
{\rm tr~}F_{\m\n}e^{-(t-t_1)\S}F_{\m\n}e^{-t_1\S} = e^{-t\s}\la\{{1\by
2}{\rm cosh}(tbw)\la(F_{\m\n}^3\ra)^2+{\rm
cosh}[(t-2t_1)bw]F_{\m\n}^+F_{\m\n}^-\ra\}.
\ee
The kinetic term for the unbroken U(1) turns out to be
\be
{1\by 8}\la({1\by g^2(x)}+{1\by g^2(y)}\ra)\int
d^4x\la(F_{\m\n}^3\ra)^2,
\ee
where $g^2$ is given in \bl{fg2} and as in section \ref{two}
$x=(\s+bw)/\L^2, y=(\s-bw)/\L^2$. For the broken generators, the
result is
\be
{1\by 2G^2(x,y)}\int d^4x F_{\m\n}^+F_{\m\n}^-
\ee
where
\be
{1\by G^2(x,y)} = {N\by 2(bw)^2}\int{d^4k\by(2\pi)^4}\la[{k^2+\s\by
2bw}{\rm ln}{k^2+\s+bw\by k^2+\s-bw}-1\ra].
\lb{g2int}
\ee
The integral over $k^2$ can be done,
\be
{1\by G^2(x,y)} = {N\by 8\pi^2(x-y)^2}\la[\la({x+y\by
4}\ra){I(x)-I(y)\by x-y} + {1\by 3}{J(x)-J(y)\by x-y} - {1\by 2}\ra],
\lb{fcg2}
\ee
where the functions $I$ and $J$ are
\bea
I(x) &=& {\rm ln}(1+x)-x^2{\rm ln}(1+1/x)+x, \nl
J(x) &=& {\rm ln}(1+x)+x^3{\rm ln}(1+1/x)+x/2-x^2.
\eea
$G^2(x,y)$ is positive for $x,y$ positive which is the region of
interest. It is a monotonically increasing function of $x+y$ in this
region. It tends to $g^2(x)$ as we approach the critical line $x=y$.
This is as it should be, since the SU(2) gauge symmetry is unbroken
along the critical line and the kinetic terms for the broken and
unbroken generators should add up to form SU(2) invariant kinetic
terms.

The gauge bosons of the broken generators receive mass terms. This
arises from Eq. \bl{kts}. Taking $\S$ to be space-time independent and
writing
\be
A_\m = {1\by\sqrt{2}}A_\m^+\s_+ + {1\by\sqrt{2}}A_\m^-\s_- + {1\by
2}A_\m^3\s_3,
\ee
one first finds
\be
{\rm tr~}{\cal D}_\m e^{-(t-t_1)\S}{\cal D}_\m e^{-t_1\S} =
4e^{-t\s}{\rm sinh}[(t-t_1)bw]{\rm sinh}(t_1bw)A_\m^+A_\m^-.
\lb{dds}
\ee
This yields for the mass terms
\be
{\cal V}^2(x,y)\int d^4x A_\m^+A_\m^-
\ee
where
\be
{\cal V}^2(x,y) =
N\int{d^4k\by(2\pi)^4}\la[{k^2+\s\by(k^2+\s)^2-(bw)^2}-{1\by 2bw}{\rm
ln}{k^2+\s+bw\by k^2+\s-bw}\ra].
\ee
Here too, the integral over $k^2$ can be done, but it is instructive
to rewrite the result in terms of $G^2(x,y)$. Note that this integral
can be obtained by differentiating the one in Eq. \bl{g2int} with
respect to $\s$. This results in
\be
{\cal V}^2(x,y) = -{1\by 2}\L^2(x-y)^2\la({\pa\by\pa x}+{\pa\by\pa
y}\ra)\la[{1\by G^2(x,y)}\ra].
\lb{fv2}
\ee
It then follows that ${\cal V}^2(x,y)$ is positive for $x,y$ positive,
the region we are interested in. As we approach the critical line
$x=y$,
\be
{\cal V}^2(x,y) \to -{1\by 2}\L^2(x-y)^2{d\by dx}\la({1\by
g^2(x)}\ra).
\ee
The mass-squared for $A_\m^{\pm}$ is given by
\bea
M^2 &=& G^2(x,y){\cal V}^2(x,y) \nl
&\to& {1\by 2}\L^2(x-y)^2{d\by dx}{\rm ln}g^2(x) \sp {\rm
as~}x\to y.
\eea
It is thus expressible in terms of the running coupling constant close
to the critical line and vanishes along the critical line as expected.

For the completely broken phase $\S={\rm diag}(x,0)\L^2$ and
$ZZ^\dag={\rm diag}(0,v^2)$, we still have the above results but with
$y=0$. The induced kinetic term for the U(1) field now suffers from an
infrared divergence and suggests introducing an infrared cutoff $\m$
for $y$. The mass squared result obtained above, though relevant with
$y=0$, now receives an additional contribution from the kinetic terms
for the $Z$ fields:
\be
\b N{\rm tr}\la(A_\m^2ZZ^\dag\ra) = {1\by 2}\b Nv^2A_\m^+A_\m^- +
{1\by 4}\b Nv^2\la(A_\m^3\ra)^2.
\ee
This makes the U(1) field $A_\m$ massive with a mass squared
$\approx\b Nv^2g^2(\m)/2$. The additional contribution to mass squared
for $A_\m^{\pm}$ is $\b Nv^2G^2(x,0)/2$.

\subsection{The case of SO(10)}

The computations for SO(10) are along the same lines. Consider the
SU(5)$\times$U(1) phase. Here $\S={\rm diag}(x,y,z)\L^2$ along 10,
$\bar{5}$ and 1, where as in section \ref{two} $x=(3\s+\r+bw)/\L^2,
y=(2\s+9\r-3bw)/\L^2, z=(25\r+5bw)/\L^2$. One then easily computes the
kinetic terms for the SU(5) generators:
\be
{1\by 4}\la[{3\by g^2(x)}+{1\by g^2(y)}\ra]\int
d^4x\la(F_{\m\n}^A\ra)^2
\lb{ec1}
\ee
where $A$ runs over the 24 generators. The function $g^2$ has been
defined earlier in Eq. \bl{fg2}. The part involving $g^2(x)$ arises
from the 10, while the one involving $g^2(y)$ from the $\bar{5}$.
Factor 3 is a consequence of the fact that ${\rm tr}(T_aT_b)$ for the
24 $T$'s of SU(5) is 3 times in the 10 as in the $\bar{5}$. For the
broken generators along the 10 and $\bar{10}$ appearing in the
decomposition $45=24+10+\bar{10}+1$, we have
\be
{1\by 2}\la[{3\by G^2(x,y)}+{1\by G^2(x,z)}\ra]\int
d^4x\abs{F_{\m\n}^a}^2,
\lb{ec2}
\ee
where $a$ runs over the 10 generators and $G^2$ is the same function
defined earlier in Eq. \bl{fcg2}. The kinetic term for the U(1) field
is
\be
{1\by 16}\la[{2\by g^2(x)}+{9\by g^2(y)}+{5\by g^2(z)}\ra]\int
d^4x\la(F_{\m\n}^{45}\ra)^2,
\lb{u1f}
\ee
where the superscript 45 denotes the U(1) direction. Note that as we
approach the critical line $x=y=z$, gauge symmetry breaking
disappears, $G^2\to g^2$ and Eqs. \bl{ec1}, \bl{ec2} and
\bl{u1f} add up to form SO(10) invariant kinetic terms as expected.

It is straightforward to compute the mass terms for the 10 and
$\bar{10}$ gauge bosons. First we note that
\bea
{\rm tr~}{\cal D}_\m e^{-(t-t_1)\S}{\cal D}_\m e^{-t_1\S} =
4&&\la\{3e^{-t(x+y)/2}{\rm sinh}[(t-t_1)(x-y)/2]{\rm sinh}[t_1(x-y)/2]
+ \ra. \nl
&&\la.e^{-t(x+z)/2}{\rm sinh}[(t-t_1)(x-z)/2]{\rm
sinh}[t_1(x-z)/2]\ra\}\abs{A_\m^a}^2,
\eea
where a $\L^2$ has been absorbed into the $t$'s in the r.h.s.
Comparing this with Eq. \bl{dds}, we get for the mass terms
\be
\la[3{\cal V}^2(x,y) + {\cal V}^2(x,z)\ra]\abs{A_\m^a}^2,
\ee
where ${\cal V}^2$ is the same function defined earlier in Eq.
\bl{fv2}. This gives the mass-squared
\be
M^2 = \la[{3\by G^2(x,y)}+{1\by G^2(x,z)}\ra]^{-1}\la[3{\cal
V}^2(x,y) + {\cal V}^2(x,z)\ra]
\ee
for the 10's. As $x,y$ and $z$ tend to be the same,
\be
M^2 \to {1\by 8}\L^2\la[3(x-y)^2+(x-z)^2\ra]{d\by dx}{\rm
ln}g^2(x).
\ee
Here too, one obtains an expression in terms of the running coupling
constant close to the critical line. It vanishes along the critical
line as expected.

For the SU(5) phase $\S={\rm diag}(x,y,0)\L^2$ and $ZZ^\dag={\rm
diag}(0,0,v^2)$, the above results are still relevant but with $z=0$.
As in the case of SU(2), the induced kinetic term for the U(1) field
suffers from an infrared divergence and suggests introducing an
infrared cutoff $\m$ for $z$. The mass squared result obtained above,
though relevant with $z=0$, receives an additional contribution from
the kinetic terms for the $Z$ fields:
\be
\b N{\rm tr}\la(A_\m^2ZZ^\dag\ra) = {1\by 2}\b Nv^2\abs{A_\m^a}^2 +
{5\by 8}\b Nv^2\la(A_\m^{45}\ra)^2.
\ee
This makes the U(1) field massive with a mass squared $\approx\b
Nv^2g^2(\m)/2$. The additional contribution to the mass squared for
the 10's is
\be
{1\by 2}\b Nv^2\la[{3\by G^2(x,y)}+{1\by G^2(x,0)}\ra]^{-1}.
\ee

\section{Global Symmetry and the Goldstone Modes}
\lb{glo}

All the models we discussed have a global U($N$) symmetry. In addition
to the gauge symmetry, this global symmetry could also suffer
breakdown. It remains to investigate this breaking and the resulting
Goldstone bosons and other massless particles if any.

In the Grassmannian model, of the two phases, the unbroken phase
retains this global symmetry. In this phase, only $\S$ gets
expectation value; but $\S$ is a singlet under the global symmetry.
$\S$ expectation value however gives mass to all the $Z$ scalars. The
model has no massless particles in this phase. In the broken phase,
$Z$ gets an expectation value breaking the global symmetry in addition
to the gauge symmetry. $\S$ expectation value is now zero and all of
the $2NM$ real components of $Z$ are hence massless. Some of these are
the would-be Goldstone bosons, eaten away by the broken gauge
generators. Because the gauge symmetry U($M$) is completely broken
down, they are $M^2$ in number.  It turns out that all those
remaining, $2NM-M^2$ in number, are the Goldstone bosons associated
with the broken generators of the global symmetry.  There are no
unaccounted massless particles. To see this, choose the $Z$
expectation value to be of the form
\be
Z \pro \la(I_M,0_{M,N-M}\ra),
\ee
where $I_M$ is an identity matrix of order $M$ and $0_{M,N-M}$ is a
zero matrix of order $M\times(N-M)$. This expectation value breaks the
global symmetry from U($N$) down to U($N-M$). The number of broken
global generators are now easily computed; they are $2NM-M^2$ in
number. Along with the would-be Goldstone bosons, they account for all
the $2NM$ massless particles.

Coming to our type two models, the global symmetry is broken down when
$Z$ gets an expectation value. In the SU(2)$\to$U(1) phase or the
SO(10)$\to$SU(5)$\times$U(1) phase of our examples, the global
symmetry remains unbroken. $\S$, a singlet under the global symmetry,
also picks up an expectation value in this phase making all the $Z$
scalars massive. There are no massless states. In the other
interesting phase of our examples, $Z$ picks up an expectation value
along some direction, a singlet of SU(5) in the case of SO(10). $\S$
expectation value in that direction is forced to zero.  This will
introduce $2N$ real massless states of $Z$. The $Z$ expectation value
can be arranged to be of the form
\be
Z \pro \la(v,0_{M,N-1}\ra),
\ee
where $v$ is a column vector pointing in the singlet direction. This
implies that the global symmetry is broken down from U($N$) to
U($N-1$). There are $2N-1$ Goldstone bosons associated with this
breaking. The remaining one massless state of $Z$ is a would-be
Goldstone boson eaten away by the broken gauge generator. This is
consistent with the fact that $Z$ expectation value of the above type
breaks one additional gauge generator. Again, there are no unaccounted
massless states.

\section{Discussions and conclusions}
\lb{dis}

We have presented an approach to composite gauge bosons that allows
for partially broken gauge symmetries. It is a generalization of the
well-known Grassmannian models that otherwise allow for either
unbroken or completely broken gauge symmetries.  In our approach, it
is also possible to incorporate interesting potential terms leading to
a rich phase structure. Even the simplest model based on SU(2) is not
amenable to analytical handling of its phases; numerical investigation
is called for. For models that are physically interesting in
connection with unified theories, even a numerical analysis of all the
phases is a challenging endeavor.

We have illustrated our approach with an SU(2) example and analyzed in
some detail an SO(10) example that could be of interest to unified
models.  What is remarkable of this exercise is that a set of
equations governed by only two parameters gives rise to a rich set of
solutions with interesting symmetry breaking patterns. There exist
regions of the parameter space where SU(2) breaks down to U(1). In the
case of SO(10), symmetry breaking to SU(5) or to SU(5)$\times$U(1) or
perhaps to some other subgroups is possible. These examples help
realize our goal of constructing an induced gauge theory with
composite gauge bosons having partial symmetry breaking.

We have computed the properties of the composite fields, the gauge
bosons and the Higgs scalars, by doing a derivative expansion of the
large $N$ effective action. Because we need an expansion that does not
perturb the Higgs field, we cannot utilize the canonical expansions
available in the literature. We have developed a suitable derivative
expansion in the Appendix and have used it to compute the kinetic
terms and the mass terms for the composites in the various phases.

We have not addressed the issue of renormalizability of Grassmannian
models or our generalized ones. It is interesting to note that the
theory at large $N$ exhibits a critical point which extends to a
critical line in the presence of a potential. It is known that the
critical points or lines can, and in many cases do, soften the
ultraviolet behavior. This softening is probably not sufficient enough
to help renormalize the theory in four dimensions and one may have to
include other relevant operators in the Lagrangian. In this
connection, we note that certain four dimensional Grassmannian models
of composite gauge fields have been studied on the lattice and shown
to be renormalizable\fr{pal}. Their phase structures and their
relation to continuum theories remain unexplored.

This work is supported in part by U. S. Department of Energy, Grant
No. DEFG-ER91-40672.

\br
\rf{old}
J. D. Bjorken, Ann. Phys. (N.Y.) 24, 174 (1963); G. S. Guralnik, \pr
136, 1404 (1964); T. Eguchi, \prd 14, 2755 (1976); H. Terezawa, Y.
Chikashige and K. Akama, {\it ibid.} 15, 480 (1977); C. Bender, F.
Cooper and G. Guralnik, Ann. Phys. (N.Y.) 109, 165 (1977); K. Shizuya,
\prd 21, 2237 (1980). D. Amati, R. Barbieri, A. C. Davis and G.
Veneziano, \plb 102, 408 (1981); A. Hasenfratz and P. Hasenfratz, \plb
297, 166 (1992).
\rf{cpn}
H. Eichenherr, \npb 146, 215 (1978); A. D'Adda, P. DiVecchia and M.
Lusher, {\it ibid.} 146, 63 (1978); 152, 125 (1979); E. Witten, {\it
ibid.} 149, 285 (1979).
\rf{gms}
E. Gava, R. Jengo and C. Omero, \plb 81, 187 (1979); \npb 158, 381
(1979); E. Bre\'{z}in, S. Hikami and J. Zinn-Justin, {\it ibid.} 165,
528 (1980); S. Duane, {\it ibid.} 168, 32 (1980). We do not include
explicit symmetry breaking terms as in Bre\'{z}in et. al. Further, our
discussion is carried out in four dimensional Euclidean space-time.
\rf{pal}
A generalization on the lattice has been studied by F. Palumbo in \prd
48, 1917 (1993) with bosonic or fermionic constituents. Its phase
structure has not been investigated. The generalization we introduce
is not confined to any specific regularization, and our remarks are
applicable to models whose phase structure is known at least at large
$N$.
\rf{our}
B. S. Balakrishna and K. T. Mahanthappa, \prd 49, 2653 (1994).
\rf{hig}
A negative $b$ corresponds to the adjoint scalar $W$ having a negative
mass squared in its potential, as in the Higgs mechanism.
\er

\appendix

\section{Derivative Expansion}
\lb{der}

\def\pg{
\begin{picture}(18,6)(0,0)
\put(6,3){\oval(6,4)[t]}
\put(12,3){\oval(6,4)[b]}
\end{picture}
}

\def\pgi{
\begin{picture}(18,6)(0,0)
\put(3,3){\circle*{3}}
\put(6,3){\oval(6,4)[t]}
\put(12,3){\oval(6,4)[b]}
\end{picture}
}

\def\pgf{
\begin{picture}(18,6)(0,0)
\put(6,3){\oval(6,4)[t]}
\put(12,3){\oval(6,4)[b]}
\put(15,3){\circle*{3}}
\end{picture}
}

In this appendix, we carry out a derivative expansion of the effective
action
\bea
S_{eff} &=& -N{\rm ~Tr~ln}\la(-D^2+\S\ra)\nl
&=& -N\int_0^{\inf}{dt\by t}{\rm Tr}\la(e^{tD^2-t\S}\ra),
\eea
where `Tr' represents an integral over space-time and a trace over the
internal indices. In the second step above, we have used the Schwinger
representation. It is an equivalent representation at the level of
equations of motion and at all orders in the derivative expansion
except the lowest one. The lowest order yielding the effective
potential is handled separately in the paper. The Schwinger
representation involves an exponential rather than the logarithm and
is hence better suited for analysis. We thus have to compute the
`trace'
\bea
{\rm Tr}\la(e^{tD^2-t\S}\ra) &=& \int
d^4x{d^4k\by(2\pi)^4}e^{-ikx}{\rm tr~}e^{tD^2-t\S}\la(e^{ikx}\ra)\nl
&=& \int d^4x{d^4k\by(2\pi)^4}{\rm tr~}e^{t(ik+D)^2-t\S}(1)\nl
&=& \int d^4x{d^4k\by(2\pi)^4}e^{-tk^2}{\rm tr~}e^{-t\S+t(2ikD+D^2)}(1),
\eea
where `tr' is a trace over the internal indices. One now expands the
exponential inside the trace and computes different terms to obtain a
series representation for the effective action. In the literature, to
our knowledge, one expands the $\S$ term as well. This is not suited
for our purpose as we intend to keep all orders in $\S$. This suggests
that we do a perturbation theory in $2ikD+D^2$ alone. To this end, we
use the following result due to Feynman:
\be
e^{-t(H+V)} = \sum_{n=0}^{\inf}\int_0^tdt_1\int_0^{t_1}dt_2\cdots
dt_ne^{-(t-t_1)H}(-V)e^{-(t_1-t_2)H}(-V)\cdots(-V)e^{-t_nH}.
\lb{fey}
\ee
In our case $H=\S$ and $-V=2ikD+D^2$. This leads to an expansion in
$2ikD+D^2$. Rearranging the terms one obtains a derivative expansion,
that is, an expansion in $D$:
\be
S_{eff} = -N\int_0^{\inf}{dt\by t}{\rm Tr}\la(e^{tD^2-t\S}\ra) =
-N\sum_{n=0}^{\inf}\int d^4x\int{d^4k\by(2\pi)^4}\int_0^{\inf}{dt\by
t}e^{-tk^2}{\rm tr~}a_n,
\ee
where $a_n$ is of order $D^{2n}$. The calculations are quite involved.
The result to order $D^4$ is
\bea
a_0 &=& e^{-t\S}, \nl
a_1 &=& -{1\by 2}\int_0^tdt_1{\cal D}_\m
e^{-(t-t_1)\S}{\cal D}_\m e^{-t_1\S}, \nl
a_2 &=& -{1\by
2t}\int_0^tdt_1t_1(t-t_1)F_{\m\n}e^{-(t-t_1)\S}F_{\m\n}e^{-t_1\S},
\eea
where ${\cal D}{\cal O}$ for some object ${\cal O}$ is $[D,{\cal O}]$.
The result for $a_2$ is not complete. However, our interest is in its
contribution to terms quadratic in $A_\m$ with a space-time
independent $\S$ and in this respect it is complete.

A brief account of the calculations now follows. We write the
expansion \bl{fey} symbolically as
\be
e^{-t(H+V)} = \sum_{n=0}^{\inf}\pgi(-V)\pg(-V)\pg\cdots\pg(-V)\pgf~,
\ee
where again $H=\S$ and $-V=2ikD+D^2$. Symbol $\pg$ denotes a
`propagator' of the kind exp${[-(t_1-t_2)H]}$. The beginning and the
end of a term of the above kind is indicated by black dots. First we
note that only terms with an even number of $D$'s are relevant. Those
with odd number of $D$'s come with an odd number of $k$'s and their
contributions vanish after the $k-$integration. Quite often, we make
use of the following reduction to simplify results:
\be
f(t{\rm 's})\cdots{\cal O}(t_{i-1})\pg 1\pg{\cal O}(t_{i+1})\cdots =
\int_{t_i}^{t_{i-1}}d\t f(t_i\to\t)\cdots{\cal
O}(t_{i-1})\pg{\cal O}(t_{i})\cdots,
\ee
given any function $f$. In the r.h.s, $t_i$ is first absent and we
have hence replaced $t_{i+1}$ by $t_i$ and so on with $t$'s of higher
indices.  If, for instance, the function $f$ were absent or is
independent of $t_i$, this reduction introduces $t_{i-1}-t_i$ in the
r.h.s. Another property we make use of to simplify results is the
presence of an overall trace and a space-time integral that lets us
rearrange terms in some expressions.

Now we come to the calculations. At the lowest order we have
$a_0=\pg={\rm exp}(-t\S)$ giving us the effective potential. At the
next order,
\be
a_1 = \pgi 2ikD\pg 2ikD\pgf + \pgi D^2\pgf~.
\ee
The first term can be simplified,
\be
\pgi 2ikD\pg 2ikD\pgf = -4k_\m k_\n\pgi D_\m\pg D_\n\pgf =
-{2\by t}\pgi D\pg D\pgf~,
\ee
where we have replaced $k_\m k_\n$ by $\d_{\m\n}/(2t)$ as the two
would yield identical results after $k-$integration. If ${\cal D}$
represents the action $[D,{\cal O}]$ for any ${\cal O}$ immediately
next to it, one easily verifies that
\bea
\pgi D\pg D\pgf &=& \pgi {\cal D}\pg D\pgf + \pgi 1\pg D^2\pgf \nl
&=& \pgi {\cal D}\pg D\pgf + (t-t_1)\pgi D^2\pgf~.
\eea
Alternately
\bea
\pgi D\pg D\pgf &=& -\pgi D{\cal D}\pg 1\pgf + \pgi D^2\pg 1\pgf \nl
&=& -\pgi{\cal D}^2\pg 1\pgf - \pgi{\cal D}\pg D\pgf +
t_1\pgi D^2\pgf.
\eea
Adding the two results,
\be
2\pgi D\pg D\pgf = -\pgi{\cal D}^2\pg 1\pgf + t\pgi D^2\pgf~.
\ee
Putting these together, we have
\be
a_1 = {1\by t}\pgi{\cal D}^2\pg 1\pgf = {1\by
2}\pgi{\cal D}^2\pgf = -{1\by 2}{\cal D}\pg{\cal D}\pgf~.
\ee
which is the result quoted earlier. The second step follows from the
sum of
\bea
\pgi{\cal D}^2\pg 1\pgf &=& t_1\pgi{\cal D}^2\pgf \sp {\rm and} \nl
\pgi{\cal D}^2\pg 1\pgf &=& \pgi 1\pg{\cal D}^2\pgf =
(t-t_1)\pgi{\cal D}^2\pgf.
\eea

The next coefficient $a_2$ can be computed along similar lines. The
result quoted earlier is obtained by keeping only terms quadratic in
$A_\m$ with a space-time independent $\S$. This simplifies the
calculations. The leftmost and rightmost $D_\m$'s get replaced by
$-iA_\m$. This gives two $A_\m$'s already so that those $D_\m$'s in
the middle get replaced by $\pa_\m$. The contributions and their
simplified results are
\bea
\pgi D^2\pg D^2\pgf &=& \pgi\pa\cdot A\pg\pa\cdot A\pgf \nl
\pgi D^2\pg 2ikD\pg 2ikD\pgf &=& -{2\by t}(t_1-t_2)\pgi\pa\cdot
A\pg\pa\cdot A\pgf \nl
\pgi 2ikD\pg D^2\pg 2ikD\pgf &=& -{2\by t}(t_1-t_2)\pgi\pa_\m
A_\n\pg\pa_\m A_\n\pgf \nl
\pgi 2ikD\pg 2ikD\pg D^2\pgf &=& -{2\by t}(t_1-t_2)\pgi\pa\cdot
A\pg\pa\cdot A\pgf \nl
\pgi 2ikD\pg 2ikD\pg 2ikD\pg 2ikD\pgf &=& 16k_\m k_\n k_\r k_\s\pgi
A_\m\pg i\pa_\n\pg i\pa_\r\pg A_\s\pgf.
\eea
Here and in the following, a $\pa$ immediately left to an $A$ as in
$\pa\cdot A$ or $\pa_\m A_\n$ acts only on that A, that is, $\pa\cdot
A=\pa_\m(A_\m)$ for instance. Note that the replacement
\be
k_\m k_\n k_\r k_\s \to {1\by
4t^2}\la(\d_{\m\n}\d_{\r\s}+\d_{\m\r}\d_{\n\s}+\d_{\m\s}\d_{\n\r}\ra)
\ee
simplifies the last contribution to
\be
{2\by t^2}(t_1-t_2)^2\la(2\pgi\pa\cdot A\pg\pa\cdot A\pgf + \pgi\pa_\m
A_\n\pg\pa_\m A_\n\pgf\ra).
\ee
Adding all the contributions, one gets
\be
\la[1-{2\by t}(t_1-t_2)\ra]^2\pgi\pa\cdot A\pg\pa\cdot A\pgf - {2\by
t}(t_1-t_2)\la[1-{1\by t}(t_1-t_2)\ra]\pgi\pa_\m A_\n\pg\pa_\m
A_\n\pgf.
\ee
Further simplification is possible due to
\be
f(t_1-t_2)\pgi{\cal O}\pg{\cal O}\pgf = t_1f(t-t_1){\cal O}\pg{\cal
O}\pgf
\ee
and
\be
f(t_1-t_2)\pgi{\cal O}\pg{\cal O}\pgf = (t-t_1)f(t_1)\pgi{\cal O}\pg{\cal
O} = (t-t_1)f(t_1){\cal O}\pg{\cal O}\pgf.
\ee
given any object ${\cal O}$. In our case $f(t-t_1)=f(t_1)$ so that
adding and dividing by 2, we get
\be
f(t_1-t_2)\pgi{\cal O}\pg{\cal O}\pgf = {t\by 2}f(t_1){\cal O}\pg{\cal
O}\pgf.
\ee
This simplifies the total contribution to
\be
{1\by 2t}(t-2t_1)^2\pa\cdot A\pg\pa\cdot A\pgf - {1\by
t}t_1(t-t_1)\pa_\m A_\n\pg\pa_\m A_\n\pgf.
\ee
Note that if one were to work in the Lorentz gauge $\pa\cdot A=0$ the
first term will vanish and the kinetic terms for the gauge fields will
arise from the second term. We will regard the second term to be a
part of the following gauge invariant combination:
\be
a_2 = -{1\by
2t}\int_0^tdt_1t_1(t-t_1)F_{\m\n}e^{-(t-t_1)\S}F_{\m\n}e^{-t_1\S}.
\ee
However, the $(\pa\cdot A)^2$ term that this generates does not agree
with what we obtained. This is to be expected since there are other
gauge invariant combinations, for instance ${\cal
D}^2e^{-(t-t_1)\S}{\cal D}^2e^{-t_1\S}$, that could generate them.

\end{document}